\documentclass[fleqn]{article}
\usepackage{graphics}

\usepackage{amsfonts}

\def\onecol{\onecolumn \mathindent 2em}

\def\noi{\noindent}

\makeatletter
\renewcommand{\section}{\@startsection{section}{1}{0pt}%
        {-3.5ex plus -1ex minus -.2ex}{2.3ex plus .2ex}%
        {\large\bf\protect\raggedright}}

\renewcommand{\subsection}{\@startsection{subsection}{2}{0pt}%
        {-3ex plus -1ex minus -.2ex}{1.4ex plus .2ex}%
        {\normalsize\bf\protect\raggedright}}

\renewcommand{\thesubsubsection}%
        {\arabic{section}.\arabic{subsection}.\arabic{subsubsection}.}

\renewcommand{\@oddhead}{\raisebox{0pt}[\headheight][0pt]{%
   \vbox{\hbox to\textwidth{\rightmark \hfil \rm \thepage \strut}\hrule}}}
\renewcommand{\@evenhead}{\raisebox{0pt}[\headheight][0pt]{%
   \vbox{\hbox to\textwidth{\thepage \hfil \leftmark \strut}\hrule}}}

\makeatother

\newcommand{\Title}[1]{\noi {\Large #1} \\}

\newcommand{\Authors}[4]{\noi
        {\large\bf #1\dag\ #2\ddag}\medskip\begin{description}
        \item[\dag]{\it #3} \item[\ddag]{\it #4}\end{description}}

\newcommand{\Abstract}[1]{\vskip 2mm \begin{center}
        \parbox{16.4cm}{\small\noi #1} \end{center}\medskip}

\newcommand{\foom}[1]{\protect\footnotemark[#1]}

\newcommand{\email}[2]{\footnotetext[#1]{e-mail: #2}
        \addtocounter{footnote}{1}}

\newcommand{\Theorem}[2]{\medskip\noi {\bf #1. \ }{\sl #2}\medskip}
\newcommand{\Ref}[1]{Ref.\,\cite{#1}}
\newcommand{\sect}[1]{Sec.\,#1}

\def\nqq{\hspace*{-2em}}
\def\nhq{\hspace*{-0.5em}}

\def\cm{\hspace*{1cm}}

\def\nx{\vspace*{-1ex}}

\def\fig{Fig.\,}

\def\Jl#1#2{{\it #1\/} {\bf #2},\ }

\def\CQG#1 {\Jl{Clas. Qu. Grav.}{#1}}
\def\DAN#1 {\Jl{Dokl. AN SSSR}{#1}}
\def\GC#1 {\Jl{Grav. \& Cosmol.}{#1}}
\def\GRG#1 {\Jl{Gen. Rel. Grav.}{#1}}
\def\JETF#1 {\Jl{Zh. Eksp. Teor. Fiz.}{#1}}
\def\JETP#1 {\Jl{Sov. Phys. JETP}{#1}}
\def\JHEP#1 {\Jl{JHEP}{#1}}
\def\JMP#1 {\Jl{J. Math. Phys.}{#1}}
\def\NPB#1 {\Jl{Nucl. Phys.}{B\ #1}}
\def\PLA#1 {\Jl{Phys. Lett.}{#1A}}
\def\PLB#1 {\Jl{Phys. Lett.}{#1B}}
\def\PRD#1 {\Jl{Phys. Rev.}{D\ #1}}
\def\PRL#1 {\Jl{Phys. Rev. Lett.}{#1}}

\newcommand{\eqsection}{\makeatletter
    \@addtoreset{equation}{section}
    \renewcommand{\theequation}{\arabic{section}.\arabic{equation}}
    \makeatother}
\def\al{&\nhq}
\def\lal{&&\nqq {}}
\def\eq{Eq.\,}
\def\eqs{Eqs.\,}
\def\beq{\begin{equation}}
\def\eeq{\end{equation}}
\def\bear{\begin{eqnarray}}
\def\bearr{\begin{eqnarray} \lal}
\def\ear{\end{eqnarray}}
\def\earn{\nonumber \end{eqnarray}}
\def\nn{\nonumber\\ {}}

\def\eql{\al =\al}

\def\dst{\displaystyle}
\def\tst{\textstyle}
\def\fracd#1#2{{\dst\frac{#1}{#2}}}
\def\fract#1#2{{\tst\frac{#1}{#2}}}
\def\Half{{\fracd{1}{2}}}
\def\half{{\fract{1}{2}}}

\def\eqdef{\stackrel{\rm def}=}
\def\e{{\,\rm e}}
\def\d{\partial}

\def\const{{\rm const}}

\def\DAL{\mathop{\raisebox{3.5pt}{\large\fbox{}}}\nolimits}

\def\MN{^{\mu\nu}}
\def\mN{_\mu^\nu}

\def\R{{\mathbb R}}
\def\S{{\mathbb S}}

\begin{document}
\onecol
 \thispagestyle{plain}

\Title{\bf Global Monopole in General Relativity}

\Authors{Kirill A. Bronnikov\foom 1}
    {,\ Boris E. Meierovich\foom 2 and Evgeny R. Podolyak}
    {Center for Gravitation and Fundamental Metrology, VNIIMS,
        3-1 M. Ulyanovoy St., Moscow 117313, Russia;\\
     Institute of Gravitation and Cosmology, PFUR,
        6 Miklukho-Maklaya St., Moscow 117198, Russia}
    {P.L. Kapitza Institute of Physics Problems,
        2 Kosygina St., Moscow 117334, Russia}

\Abstract {We consider the gravitational properties of a global
monopole on the basis of the simplest Higgs scalar triplet model
in general relativity. We begin with establishing some common
features of hedgehog-type solutions with a regular center,
independent of the choice of the symmetry-breaking potential. There are
six types of qualitative behavior of the solutions; we show, in
particular, that the metric can contain at most one simple horizon. For
the standard Mexican hat potential, the previously known
properties of the solutions are confirmed and some new results are
obtained. Thus, we show analytically that solutions with
monotonically growing Higgs field and finite energy in the static
region exist only in the interval $1<\gamma <3$, $\gamma $ being
the squared energy of spontaneous symmetry breaking in Planck
units. The cosmological properties of these globally regular
solutions apparently favor the idea that the standard Big Bang
might be replaced with a nonsingular static core and a horizon
appearing as a result of some symmetry-breaking phase transition
on the Planck energy scale.

\quad In addition to the monotonic solutions, we present and analyze a
sequence of families of new solutions with oscillating Higgs field. These
families are parametrized by $n$, the number of knots of the Higgs field,
and exist for $\gamma < \gamma_n = 6/[(2n+1) (n+2)]$; all such solutions
possess a horizon and a singularity beyond it.
}

\email 1 {kb@rgs.mccme.ru}
\email 2 {meierovich@yahoo.com; \
    http://geocities.com/meierovich }

\section{Introduction}

    According to the Standard cosmological model \cite{Zeldovich}, the
    Universe has been expanding and cooling from a split second after the
    Big Bang to the present moment and remained uniform and isotropic
    overall in doing so. In the process of its evolution, the Universe has
    experienced a chain of phase transitions with spontaneous symmetry
    breaking, including Grand Unification, electroweak phase transition,
    formation of neutrons and protons from quarks, recombination, and so
    forth. Regions with spontaneously broken symmetry, which are more than
    the correlation length apart, are statistically independent. At
    interfaces between these regions, so-called topological defects
    necessarily arise. A systematic exposition of the potential role of
    topological defects in our Universe is provided by Vilenkin and Shellard
    \cite{Vilenkin and Shellard}. The particular types of defects: domain
    walls, strings, monopoles, or textures are determined by the topological
    properties of vacuum \cite{Kibble}. If the vacuum manifold after the
    breakdown is not shrinkable to a point, then the Polyakov-t'Hooft
    monopole-type solutions \cite{Polyakov,Hooft} appear in
    quantum field theory.

    Spontaneous symmetry breaking (SSB) plays a fundamental role in modern
    attempts to construct particle theories. A symmetry in this context is
    not necessarily associated with space-time transformations. It can be a
    kind of ``internal'' symmetry as well, such as the Grand Unification
    symmetry, the electroweak and the isotopic symmetry, or even
    supersymmetry, whose transformations mix bosons and fermions.
    Topological defects, caused by spontaneous breaking of internal
    symmetries (independent of space-time coordinates), are called global.

    A fundamental property of global symmetry violation is the Goldstone
    degree of freedom. In the monopole case, the term related to the
    Goldstone boson in the energy-momentum tensor decreases rather slowly
    away from the center. As a result, the total energy of a global monopole
    grows linearly with distance, in other words, diverges. Without gravity
    such a divergence is a general property of spontaneously broken global
    symmetries.  In his pioneering paper \cite{Polyakov} Polyakov mentioned
    two possibilities of avoiding this difficulty. The first one was to
    combine a monopole with a Yang-Mills field.
    This idea was independently considered by t'Hooft \cite {Hooft}.
    This, among other reasons, gave rise to numerous papers on gauge
    (magnetic) monopoles. The second possibility was to consider a bound
    monopole-antimonopole system, whose total energy would be large
    (proportional to the distance between the components) but finite.

    One more opportunity is to take into account the self-gravity of global
    monopoles, which can in principle remove the above self-energy problem.
    This is also necessary for potential astrophysical applications.
    Such a study was first performed by Barriola and Vilenkin
    \cite{Barriola and Vilenkin}
    who found that the gravitational field outside a monopole is
    characterized by a solid angle deficit proportional to the SSB energy
    scale. Harari and Lousto showed that the gravitational mass of a
    global monopole, calculated using the Tolman integral, is negative
    \cite{Harari and Lousto}.  Solutions with a horizon for supermassive
    global monopoles were found by Liebling \cite{Liebling}, who also
    confirmed the estimate  of \Ref{Sakai et al} for the upper value of the
    symmetry breaking energy compatible with a static configuration.
    The existence of de Sitter cores inside global monopoles and other
    topological defects gave rise to the idea of ``topological inflation''
    \cite{Vilenkin (Top infl),Linde, Basu and Vilenkin}.

    For global strings in flat space, the energy per unit length (without
    gravitation) also diverges with growing distance from the axis, but only
    logarithmically. However, in general relativity integration over the
    cross-section yields a finite result \cite{MeierovichGRG, Meierovich
    GC}. The gravitational interaction thus leads to self-localization of a
    global string. Will a similar effect take place for a global monopole?
    An attempt to answer this question, which appears not to be answered in
    the existing papers, was one of motivations for reconsidering the
    gravitational properties of a global monopole.

    The previous studies have used the boundary condition according to which
    the symmetry-breaking potential should vanish at spatial infinity.
    Our approach is different: we do not even assume the existence of a
    spatial asymptotic but require regularity at the center and try to
    observe the properties of the whole set of global monopole solutions.
    In doind so, among other quantities, we discuss the behavior of the
    total scalar field energy, which turns out to be finite in static
    regions of supermassive global monopoles.

    In \sect 2 we present the complete sets of equations for a
    static spherically symmetric gravitating global monopole in two
    most convenient coordinate systems, namely, with quasiglobal and
    harmonic radial coordinates. The general properties of static global
    monopoles are summarized in \sect 3. In \sect 4 we analyze
    analytically and numerically the specific features of a global monopole
    in the particular case of the ``Mexican hat'' potential. \sect 5
    contains a general discussion of our results, including their possible
    cosmological interpretation.

\section{Equations and boundary conditions}

\subsection{General problem setting}

    We begin with the most general form of a static, spherically symmetric
    metric, without specifying the radial coordinate $x^1=u$:
\beq
    ds^2 = g_{\mu \nu}dx^{\mu }dx^{\nu }                        \label{ds}
       =\e^{2F_{0}}dt^2
        -\e^{2F_1} du^2 - \e^{2F_\Omega} d\Omega^2.
                        \label{metric general form}
\eeq
    Here $d\Omega^2= d\theta^2 + \sin ^2 \theta d\varphi^2$ is the linear
    element on a unit sphere and $F_{0}$, $F_1$, and $F_{\Omega}$ are
    functions of $u$.

    The nonzero components of the Ricci tensor are
    (the prime denotes $d/du$)
\bear
  R_0^{\ 0} \eql \e^{-2F_1}[ F_0''
             + F_0'(-F_1' + 2F'_\Omega +F_0') ] ;
\nn
     R_1^{{ \ }1}
            \eql \e^{-2F_1}[ F_0'' + 2F''_\Omega               \label{Rmn}
        +2F'_\Omega{}^2 + F'_0{}^2
        -F_1'(2F'_\Omega + F'_0) ] ;        \label{Ricci tensor general}
\nn
    R_2^{\ 2} \eql R_3^{\ 3} = -\e^{-2F_{\Omega}}
    +\e^{-2F_1}[ F''_{\Omega} + F'_{\Omega}(-F_1'+2F_{\Omega }'+F_0') ].
\ear

    Consider the Lagrangian describing a triplet of real scalar fields
    $\phi^a$ ($a=1,2,3$) in general relativity:
\beq                                                          \label{L}
    L = \frac{R}{16\pi G} + \Half g\MN \d_\mu\phi^a \d_\nu\phi^a -V(\phi),
\eeq
    where $R$ is the scalar curvature,
    $V(\phi)$ is a potential depending on
    $\phi = \pm\sqrt{\phi^a\phi^a}$ and $G$ is the gravitational constant.
    We use the natural units such that
\beq
          \hbar =c=1,                                       \label{Units}
\eeq
    so that $G= m_{\rm pl}^{-2}$, where $m_{\rm pl}=1.22\times 10^{19}$ GeV
    is the Planck mass.

    To obtain a global monopole with unit topological charge
    \cite{Vilenkin and Shellard}, let us assume that the metric has the form
    (\ref{ds}) while $\phi^a$ comprise the following ``hedgehog''
    configuration:
\beq                                                          \label{hog}
    \phi^a = \phi(u) n^a, \cm n^a =
                \left\{
                \sin\theta \cos\varphi,\quad
                \sin\theta \sin\varphi,\quad
                \cos\theta
                \right\}.
\eeq

    The Einstein equations can be written in the form
\beq
    R\mN = -8\pi G \tilde T\mN                                \label{EE}
         = -8\pi G (T\mN - \half \delta\mN T^\alpha_\alpha)
\eeq
    where $T\mN$ is the energy-momentum tensor and
    the nonzero components of $\tilde T\mN$ are
\beq
    \tilde T_0^0 = -V, \cm                                    \label{Tmn}
    \tilde T_1^1 = -V - \e^{-2F_1} \phi'{}^2, \cm
    \tilde T_2^2 = \tilde T_3^3 = -V - \e^{-2F_\Omega} \phi^2
\eeq

    The regular center conditions for the metric (\ref{ds}) are that at
    the corresponding value $u_c$ of the coordinate $x^1=u$
\beq
    \e^{F_\Omega}\to 0; \cm   F_0 = F_{0\,c} + O (\e^{2F_\Omega}),
    \cm \e^{-F_1+F_\Omega} |F'_\Omega| \to 1.                 \label{cent}
\eeq
    The last condition is necessary for local
    flatness and provides the correct circumference
    to radius ratio for coordinate circles at small $r = \e^{F_\Omega}$.

    The scalar field energy, defined as the partial time derivative of the
    scalar field action, $E=-\d S/\d t$, is a conserved quantity for
    our static system:
\beq
    E = \int \sqrt{-g} T_0^0 d^3 x                           \label{E}
       = 4\pi \int \e^{F_0+F_1+2F_\Omega}
    \biggl[\Half \e^{-2F_1}\phi'{}^2 + \e^{-2F_\Omega}\phi^2 +V \biggr] du,
\eeq
    where $g$ is the metric tensor determinant.

    In what follows we will make some general inferences without specifying
    the potential $V(\phi)$ and then perform a more detailed study for the
    simplest and most frequently used symmetry-breaking potential
\beq
       V(\phi) = \frac 14 \lambda (\phi^a \phi^a - \eta^2)^2
               = \frac 14 \eta^4 \lambda (f^2 - 1)^2,        \label{hat}
\eeq
    where $\eta > 0$ characterizes the energy of symmetry breaking,
    $\lambda $ is a dimensionless constant and $f(u)= \phi(u)/\eta$ is the
    normalized field magnitude playing the role of an order parameter. The
    model has a global $SO(3)$ symmetry, which can be spontaneously broken
    to $SO(2)$ due to the potential wells ($V=0$) at $f=\pm 1$.

    Let us now write down the Einstein equations and the boundary
    conditions explicitly in two coordinate frames to be used.

\subsection {The quasiglobal coordinate $\rho$}

    The first choice is the coordinate $u=\rho$ specified by the condition
    $F_0 + F_1=0$. Denoting $\e^{2F_0} = \e^{-2F_1} = A(\rho)$ and
    $\e^{F_\Omega}= r(\rho)$, we obtain the metric in the form
\beq
                                                           \label{ds-rho}
    ds^2 = A(\rho)dt^2 - \frac{d\rho^2}{A(\rho)}
                            -r^2(\rho)d\Omega^2.
\eeq

    The scalar field equation
\beq
    \DAL \phi^a + \d V/\d\phi^a =0,                          \label{phi-gen}
\eeq
    where $\DAL = \nabla^\alpha\nabla_\alpha$ is the d'Alembert operator,
    and certain combinations of the Einstein equations have the form
\bear
       (Ar^2 \phi')' -2 \phi \eql r^2 dV/d\phi;              \label{phi}
\\
              (A'r^2)' \eql - 16\pi G r^2 V;                  \label{00}
\\
              2 r''/r  \eql - 8\pi G \phi'{}^2 ;              \label{01}
\\
       A (r^2)'' - r^2 A'' \eql 2(1 - 8\pi G \phi^2);         \label{02}
\\
     A' rr' + A{r'}^2 -1 \eql
            8\pi G [\half Ar^2\phi'{}^2 - \phi^2 -r^2 V],     \label{int}
\ear
    where the prime denotes $d/d\rho$. Only three of these five equations
    are independent: the scalar field equation (\ref{phi}) follows from
    the Einstein equations, while \eq (\ref{int}) is a first integral of the
    others. Given a potential $V(\varphi)$, this is a determined set of
    equations for the unknowns $r,\ A,\ \phi$.

    This choice of the coordinates is preferable for considering Killing
    horizons, which correspond to zeros of the function $A(\rho)$, since
    such zeros are regular points of \eqs (\ref{phi})--(\ref{int});
    moreover, in a close neighborhood of a horizon, the coordinate $\rho$
    defined in this manner varies (up to a positive constant factor) like
    manifestly well-behaved Kruskal-like coordinates used for analytic
    continuation of the metric \cite{cold,br01}. Therefore one can jointly
    consider regions at both sides of a horizon in terms of $\rho$, and, in
    general, the whole range of $\rho$ can contain several horizons.  For
    this reason the coordinate $\rho$ can be called {\it quasiglobal.\/}

    The regular center conditions (\ref{cent}) are fulfilled if, near
    some value $\rho_c$ of the coordinate $\rho$,
\beq
     A(\rho) = A_c + O((\rho-\rho_c)^2), \cm
     r(\rho) \approx (\rho-\rho_c)/\sqrt{A_c}.              \label{cent-rho}
\eeq

    In regions where $A < 0$ (sometimes called T-regions), if any, the
    coordinate $\rho$ is timelike and $t$ is space-like. Changing the
    notations, $t\to x \in \R$, and introducing the proper time of a
    comoving observer in a T-region,
\beq
    \tau = \int d\rho/\sqrt{|A(\rho)|},                 \label{tau}
\eeq
    we can rewrite the metric in the form
\beq                                                        \label{ds-tau}
    ds^2 = d\tau^2 - |A(\tau)| dx^2 - r^2(\tau) d\Omega^2.
\eeq
    The space-time geometry then corresponds to a homogeneous anisotropic
    cosmological model of Kantowski-Sachs (KS) type \cite{kompa,kant}, where
    spatial sections have the topology $\R \times \S^2$.

\subsection {The harmonic coordinate $u$}

    Another convenient variable, well simplifying the form of the equations,
    is the harmonic coordinate $u$ specified by the condition \cite{br73}%
\footnote
{A cylindrical version of the harmonic radial coordinate has been used
previously in the analysis of gravitational properties of current-conducting
filaments \cite{Meierovich} and cosmic strings \cite{Meierovich and
Podolyak PR,Meierovich Uspekhi}.}
\beq
    F_1 = 2F_\Omega + F_0,                                  \label{harm}
\eeq
    so that $\DAL u =0$. Then the field equations may be written as
\bear
    \phi'' - 2 \e^{F_0+F_1} \phi \eql \e^{2F_1}dV/d\phi,   \label{phi-u}
\\
    F_0'' \eql -8\pi G\e^{2F_1} V,                         \label{00-u}
\\
    F_1'' - 2F_{\Omega}'(F'_{\Omega} + 2 F'_0)             \label{11-u}
                    \eql -8\pi G(\phi'{}^2 + \e^{2F_1}V),
\\
    F''_\Omega -\e^{2( F_0 + F_\Omega)} \eql               \label{22-u}
         -8\pi G(\phi^2\e^{2(F_0 + F_\Omega)} + \e^{2F_1} V),
\\
     - \e^{-2F_\Omega} + \e^{-2F_1} (F'_{\Omega}{}^2           \label{int-u}
                    +2F'_\Omega F'_0) \eql
    8\pi G (\half \e^{-2F_1}\phi'{}^2 -\e^{-2F_\Omega}\phi^2 -V)
\ear
    where the prime denotes $d/du$.

    It is straightforward to obtain that a regular center can only
    correspond to $u\to \pm \infty$; we choose $u\to -\infty$, where one
    should have
\beq
     \e^{F_\Omega} \sim 1/|u|, \cm
     \e^{F_0} = \sqrt{A_c} (1 + O(u^{-2})), \cm
     \e^{F_1} \sim 1/u^2,                                \label{cent-u}
\eeq
    and $A_c$ is the same as in (\ref{cent-rho}).

\section{General properties of global monopoles}

\subsection {Monopoles in Minkowski space-time}

    The Minkowski metric in the usual spherical coordinates
\beq
    ds^2 = dt^2 - dr^2 -r^2 d\Omega^2                  \label{ds-flat}
\eeq
    is a special case of (\ref{ds-rho}) with $r \equiv \rho$ and $A\equiv
    1$. The only unknown in flat space-time is $\phi(r)$, and the only
    field equation is (\ref{phi}) which takes the form
\beq
       (r^2 \phi')' - 2 \phi = r^2 dV/d\phi,              \label{phi-flat}
\eeq
    where, in particular, for the potential (\ref{hat})
    $dV/d\phi = \lambda \phi (\phi^2 - \eta^2)$. In this case the scalar
    field equation can be written in terms of $f = \phi/\eta$ as
\beq                                                       \label{f-flat}
      r^{-2} (r^2f')' - 2fr^{-2} + \lambda\eta^2 f(1-f^2) =0.
\eeq

    The energy integral (\ref{E}) takes the form
\beq
    E = 4\pi \int r^2 \biggl[
        \Half \phi'{}^2 + \frac{\phi^2}{r^2} + V \biggr] dr, \label{E-flat}
\eeq
    and for its convergence, in case $V(\phi)\geq 0$, all the three terms
    should vanish quickly enough at infinity:
\beq
      \phi=o(r^{-1/2}), \cm \phi'=o(r^{-3/2}), \cm V =o(r^{-3})\label{E-fin}
\eeq
    as $r\to\infty$. This actually means that a configuration with finite
    energy is only possible with $V(0)=0$, contrary to the symmetry breaking
    assumption according to which $V$ has minima in nonsymmetric states,
    $\phi\ne 0$. In particular, the potential (\ref{hat}) does not give rise
    to global monopoles with finite energy. A consideration of self-gravity
    of the field triad $\phi^a$ is one of the ways to overcome this
    difficulty.

    The harmonic coordinate $u$ in flat space-time is connected with $r$
    by the relation $u-u_0 = \pm 1/r$, where $u_0$ is an arbitrary constant;
    choosing the minus sign, we find that $u$ ranges from $-\infty$, which
    corresponds to the center $r=0$, to $u_0$ corresponding to spatial
    infinity.

\subsection{Solutions with constant $\phi$}

    Under the assumption $\phi=\phi_0=\const$, the corresponding value
    of the potential $V(\phi_0)=V_0$ (times $8\pi G$) plays the role of a
    cosmological constant, and the Einstein equations can be integrated
    explicitly.

    Indeed, in a region where $\phi=\const$, \eq(\ref{01}) reduces
    to $r''=0$, whence $r = \alpha\rho + r_0$, $\alpha,\ r_0 =\const$.
    It remains to find $A(r)$, and this is immediately done by
    integrating \eq (\ref{02}):
\beq
    A(r) = \frac{1-\Delta}{\alpha^2} - \frac{2GM}{r} + Cr^2,
     \cm
         \Delta = 8\pi G \phi_0^2,                             \label{A1}
\eeq
    where $M$ and $C$ are integration constants. Substituting (\ref{A1})
    into (\ref{00}), we find
\beq
      C = - 8\pi G V_0/3.
\eeq
    Thus the solution is essentially determined by the values of $\phi_0$,
    $V_0$ and $M$. One more constant, $\alpha$, reflects the freedom in
    choosing the unit of time. It should be noted that this is not a
    monopole solution.  Even if we put $M=0$, which is evidently necessary
    for regularity at $r=0$, this solution with constant $\phi\ne 0$ is
    singular at the center: for (\ref{A1}) with $M=0$ the Kretschmann scalar
    at small $r$ is ${\cal K} =
    R_{\alpha\beta\gamma\delta}R^{\alpha\beta\gamma\delta} \approx  4
    \Delta^2/r^4$.

    With respect to a global monopole, two cases of the solution (\ref{A1})
    are of interest. The case $\phi _0 \equiv  0$ describes the
    symmetric state, while the case $V_{0}=0$ gives a possible asymptotic
    behavior at spatial or temporal infinity.

    In case $\phi_0=0$ (the symmetric state), putting $M=0$ (which is
    necessary for a regular center), we arrive at the de Sitter metric
\beq
     ds^2 = \biggl(1-\frac{r^2}{r_h^2}\biggr)dt^2
            - \biggl(1-\frac{r^2}{r_h^2}\biggr)^{-1}dr^2 -r^2 d\Omega^2,
     \cm
     r_h^2 = \frac{8\pi G V_0}{3}.                             \label{deS}
\eeq
    This metric has a horizon at $r=r_h$. At $r > r_h$, outside the horizon,
    $r$ becomes a timelike coordinate, and $t$ is a spacelike one.
    Changing the notations as in (\ref{tau}), (\ref{ds-tau}), we obtain
    the metric
\beq
    ds^2=d\tau^2 - \sinh^2(\tau/r_h) dx^2
        - r_h^2 \cosh^2(\tau/r_h) d\Omega^2.       \label{deS-tau}
\eeq
    This is a KS cosmology with isotropic inflationary
    expansion at late times ($\tau\to \infty$).

    In the other case, $\phi_0\ne 0$ but $V_0=0$ (the case of broken
    symmetry, such as $\phi=\eta$ in the potential (\ref{hat})), the metric
    takes the form \cite{Barriola and Vilenkin}
\beq
    ds^2=\biggl(\frac{1-\Delta}{\alpha^2}
                    - \frac{2GM}{r}\biggr) dt^2
       -\biggl(\frac{1-\Delta}{\alpha^2}
                -\frac{2GM}{r}\biggr)^{-1} dr^2
                        - r^2 d\Omega^2,       \label{qSch}
\eeq
    where the constant $M$ has the meaning of mass in the sense
    that test particles at rest at large $r$ experience an acceleration
    equal to $-GM/r^2$ in the gravitational field (\ref{qSch}).
    Furthermore, a nonzero value of $\phi_0$ leads to a solid angle deficit
    $\Delta$ defined in (\ref{A1}) at the asymptotic $r\to \infty$
    (see \cite{Vilenkin and Shellard} for more detail) and to a linear
    divergence of the integral (\ref{E}) at large $r$.

    The general case of \eq(\ref{A1}) describes the large $r$ asymptotic of
    any solution to \eqs (\ref{phi})--(\ref{int}), provided that such an
    asymptotic exists and $\phi$ tends quickly enough to a constant value.

    For the monopoles to be studied, the metric (\ref{qSch}) gives a
    large $r$ asymptotic in the case $\Delta <1$. We will also consider
    solutions with $\Delta > 1$, for which a static asymptotic is absent.
    Then the metric (\ref{qSch}) describes cosmological evolution at late
    times.

\subsection {General properties of solutions with varying $\phi$}

    Consider now the general form of \eqs(\ref{phi})--(\ref{int}) with
    varying $\phi$, without specifying the potential $V(\phi)$.

    First of all, we note that, due to (\ref{01}), $r'' \leq 0$, which rules
    out any nonsingular configurations without a center such as wormholes
    and horns (see Theorem 1 of \Ref{br01} for more detail).

    Second, \eq(\ref{02}) can be rewritten in the form
\beq
    (r^4 B')' = -2 (1- 8\pi G \phi^2), \cm B \eqdef A/r^2,      \label{02B}
\eeq
    and at a point where $B'=0$ we have $r^4B'' = -2(1-8\pi G\phi^2)$. Hence
    it follows that, as long as $\phi^2 < 1/(8\pi G)$ (the $\phi$ field does
    not reach trans-Planckian values), $B'' <0$ at possible extrema of the
    function $B$. In other words, $B$ cannot have a regular minimum.

    Our interest is in systems with a regular center satisfying the
    conditions (\ref{cent-rho}), so that both $A(\rho)>0$ and $B(\rho)>0$
    near $\rho=\rho_c$. At a possible horizon $\rho=h$, both $A$ and $B$
    vanish, and, since it cannot be a minimum of $B$, $B < 0$ at $\rho>h$
    near the horizon. At greater $\rho$ the function $B(\rho)$, having no
    minima, can only decrease and will never return to zero; hence $A= Br^2
    < 0$ at $\rho>h$. We conclude that there can be no more than one
    horizon, and, if it exists, it is simple (corresponds to a simple zero
    of $A(\rho)$).  Since the global causal structure of space-time is
    determined (up to possible identifications of isometric hypersurfaces)
    by the number and disposition of Killing horizons
    \cite{br79,katan,strobl}, we have the following result:

\Theorem{Statement 1}
    {Under the assumption that $\phi^2 < 1/(8\pi G)$ in the whole space, our
    system with a regular center can have either no horizon, or one simple
    horizon, and in the latter case its global structure is the same as that
    of de Sitter space-time.  }

    The above reasoning is in essence the same as in the proof of Theorem 2
    of \Ref{br01} on the disposition of horizons in scalar-vacuum
    space-times. It uses only \eq(\ref{02}) where the potential $V$ does
    not enter. Therefore the conclusion is valid for systems with any
    potentials, positive or negative.

    Let us return to \eq (\ref{01}), according to which $r''\leq 0$. Since
    at a regular center $r' >0$, this leaves three possibilities for the
    function $r(\rho)$ (see Fig.\,1):

\begin{description}
\item[(a)]
    monotonic growth with a decreasing slope, but $r\to \infty$ as
    $\rho\to\infty$,

\item[(b)]
    monotonic growth with $r\to r_{\max} <\infty$ as
    $\rho\to\infty$, and

\item[(c)]
    growth up to $r_{\max}$ at some $\rho_1 < \infty$ and further
    decrease, reaching $r=0$ at some finite $\rho_2 > \rho_1$.
\end{description}

    In each case, according to Statement 1, a horizon can occur at some
    $\rho =h$ within the range of $\rho$, so that at $\rho >h$ we have a
    T-region with the geometry of a KS cosmological model.

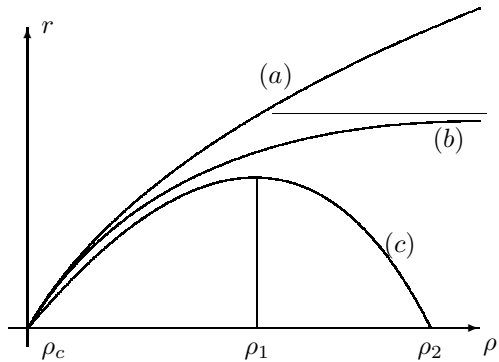
\begin{figure}
\centering
\unitlength=.5mm
\special{em:linewidth 0.4pt}
\linethickness{0.4pt}
\begin{picture}(143.00,105.00)
\put(15.00,20.00){\vector(1,0){125.00}}
\put(20.00,15.00){\vector(0,1){85.00}}
\bezier{620}(20.00,20.00)(47.00,68.00)(140.00,105.00)
\bezier{604}(20.00,20.00)(53.00,75.00)(140.00,75.00)
\bezier{776}(20.00,20.00)(87.00,100.00)(127.00,20.00)
\linethickness{0.2pt}
\put(85.00,77.00){\line(1,0){57.00}}
\put(81.00,60.00){\line(0,-1){40.00}}
\put(23.00,100.00){\makebox(0,0)[lc]{$r$}}
\put(23.00,16.00){\makebox(0,0)[lt]{$\rho_c$}}
\put(81.00,16.00){\makebox(0,0)[ct]{$\rho_1$}}
\put(127.00,16.00){\makebox(0,0)[ct]{$\rho_2$}}
\put(86.00,87.00){\makebox(0,0)[cc]{$(a)$}}
\put(132.00,70.00){\makebox(0,0)[cc]{$(b)$}}
\put(119.00,42.00){\makebox(0,0)[cc]{$(c)$}}
\put(143.00,15.00){\makebox(0,0)[cc]{$\rho$}}
\end{picture}
\caption{Possible behavior of $r(\rho)$ in global monopole solutions}
\end{figure}

    We conclude that there are six classes of qualitative behaviors of
    the solutions, i.e., (a), (b), (c), each with or without a horizon, the
    latter circumstance to be labelled with the symbol 1 or 0, respectively.
    Thus, all solutions with a spatial asymptotic belong to class (a0). Class
    (b0) includes space-times ending with a ``tube'' consisting of
    two-dimensional spheres of equal radius. Class (c0) solutions contain
    a second center at $\rho=\rho_2$, and this center can a priori be
    regular or singular. We thus obtain a static analogue of closed
    cosmologies.  Classes (a1), (b1), (c1) describe different late-time
    cosmological behaviors in the two directions corresponding to $\S^2$,
    whereas the fate of the third spatial direction ($\R$) is determined by
    the function $A(\rho)$. In particular, the possible de Sitter asymptotic
    (\ref{deS-tau}) belongs to class (a1) solutions, and in this case the
    expansion is isotropic at late times. On the other hand, class (c1)
    contains models which behave at late times like the Schwarzschild
    space-time inside the horizon, contracting to $r=0$.

    This classification is obtained without any assumptions about
    $V(\phi)$. Solutions with given $V(\phi)$ will contain some of these
    classes, not necessarily all of them.

    In case $V\geq 0$, \eq (\ref{00}) leads to one more important
    observation: since at a regular center $A'r^2 =0$, we can write
    (\ref{00}) in the integral form
\def\oro{{\bar\rho}}
\beq
      A'r^2 = -16\pi G\int_0^\rho V(\oro)\, r^2(\oro)\, d\oro.\label{00int}
\eeq
    Thus, unless $V\equiv 0$, $A(\rho)$ is a decreasing function.
    \eq(\ref{00int}) leads to the following conclusions.

\Theorem{Statement 1a}
    {If $V(\phi)\geq 0$, our system with a regular center can have either no
    horizon, or one simple horizon, and in the latter case its global
    structure is the same as that of de Sitter space-time.  }
\nx

\Theorem{Statement 2}
    {If $V(\phi)\geq 0$, the second center in class (c0) solutions is
    singular.
    }
\nx

\Theorem{Statement 3}
    {If $V(\phi)\geq 0$ and the solution is asymptotically flat, the mass
    $M$ of the global monopole is negative.
    }
\nx

    Statement 1a shows that, for nonnegative potentials, the assumption
    $\phi^2 < 1/(8\pi G)$ in Statement 1 is unnecessary, and the causal
    structure types are known for any magnitudes of $\phi$.

    Statement 2 follows from $A'(\rho_2) < 0$, whereas at a regular center
    it should be $A'=0$, see (\ref{cent-rho}). $A'(\rho_2)=0$ could only be
    possible with $V\equiv 0$, but in this case the only solution
    with a regular center is trivial (flat space, $\phi=0$).

    In Statement 3, asymptotic flatness is understood up to the solid angle
    deficit, i.e., $r=\rho$ and $A$ is given by (\ref{A1}) with $C=0$ at
    large $\rho$. Then, (\ref{00int}) for $\rho\to \infty$ gives $2GM$ on
    the left-hand side and a negative quantity on the right.

    To our knowledge, this simple conclusion, valid for all nonnegative
    potentials, has been so far obtained only numerically for the particular
    potential (\ref{hat}) \cite{Harari and Lousto}. Note that Statement 3
    is an extension to global monopoles of the so-called generalized Rosen
    theorem \cite{brsh91,br01}, previously known for scalar-vacuum
    configurations.

    Thus, even before studying particular solutions with the potential
    (\ref{hat}), we have a more or less complete knowledge of what can be
    expected from such global monopole systems.

\begin{figure}\centering
    \includegraphics{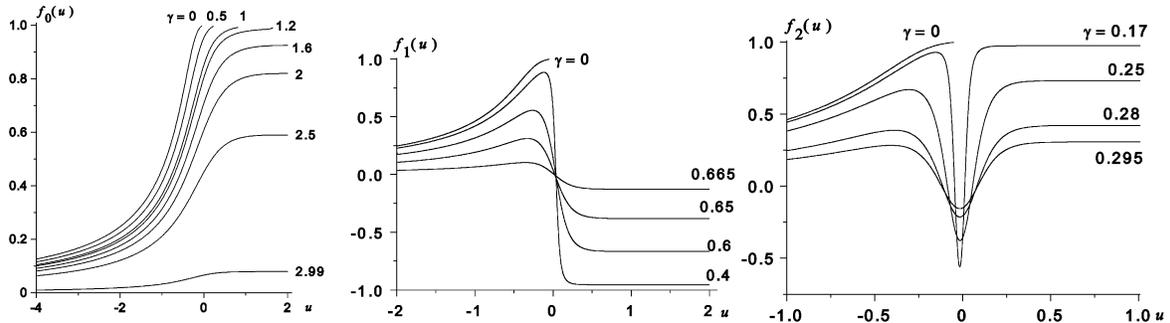}
    \caption{\label{fig:Figurea2} The field magnitude $f$ as a function of the harmonic coordinate $u$
     for different values of $\gamma$. Solutions with monotonically growing
     $f= f_{0}( u) $ (a) exist for $0<\gamma <\gamma _{0}=3.$ In the region
     $\gamma <\gamma _{1}=2/3$ there are solutions with $f=f_{1}( u) $,
     changing their sign once (b), in the region $\gamma <\gamma _{2}=0.3$
     there are solutions with $f=f_{2}( u) $ changing their sign twice (c).
     As $\gamma \to \gamma _{n}-0$, the function $f_{n}( u) $
     vanishes in the whole range $-\infty <u<\infty $ from the center to
     the horizon.}
\end{figure}

\section{Mexican hat potential}

\subsection{Equations and boundary conditions}

    Further analysis is performed for the particular ``Mexican
    hat'' potential (\ref{hat}). For numerical integration we prefer to
    use the harmonic coordinate $u$ and to work with \eqs
    (\ref{phi-u})--(\ref{22-u}). This variable enters into the equations only
    via derivatives and is thus invariant under translations $u\to
    u+\const$.

    Introducing the dimensionless quantities
\beq
     \tilde u = u/(\sqrt{\lambda}\eta),                      \label{dim-0}
  \cm
     \e^{\tilde{F}_\Omega} = \sqrt{\lambda}\eta\e^{F_\Omega},
  \cm
     \e^{\tilde{F}_1} = \lambda\eta^2 \e^{F_1}
\eeq
    we exclude the parameter $\lambda$ from the
    equations. Indeed, omitting the tildes, we obtain
\bear
     f'' \eql \e^{2 (F_0 +F_\Omega)} [ 2-\e^{2F_\Omega}(1-f^2) ] f,
                                                            \label{Eq-f}
\\
     F''_0 \eql -\frac{\gamma }{4}\e^{2(F_0+2F_\Omega)}
                    (f^{2}-1)^2,                \label{Eq-F0}
\\
    F''_\Omega \eql \e^{2 (F_0+F_\Omega)}
    [ 1-\gamma f^2-\frac{\gamma }{4}\e^{2F_\Omega} (1- f^2)^2]
                                                            \label{Eq-Fom}
\ear
    The condition (\ref{harm}) is preserved for the newly defined
    quantities, but the metric now reads
\beq                                                        \label{ds_dim0}
     ds^2 = \e^{2F_0}dt^2
            - \frac{\e^{2F_1}du^2 + \e^{2F_\Omega}d\Omega^2}{\lambda\eta^2}.
\eeq

    The boundary conditions at $u \to -\infty$ are
\beq                                                        \label{bound_c}
    f=0,        \qquad F_{0}=0,
    \qquad F'_0 =0, \qquad   F_{\Omega }=-\ln(-u) + o(1/|u|),
        \qquad {\rm as} \quad u \to -\infty ,
\eeq
    They follow from the requirement of regularity at the center and a
    particular choice of the time unit ($F_0=0$) and the origin
    of the $u$ coordinate (the fourth condition).

    There remains only one dimensionless parameter in
    \eqs(\ref{Eq-f})--(\ref{Eq-Fom}),
\beq
        \gamma =8\pi G\eta ^{2},                            \label{gamma}
\eeq
    it is the squared energy of symmetry breaking in Planck units.

    It is easy to obtain that $\gamma=1$ is a critical value of this
    parameter. Indeed, if we suppose the existence of a large $r$ asymptotic
    at which $f\to 1$, i.e., the field tends to the minimum of the potential
    (\ref{hat}), then the asymptotic form of the metric at large $r$ is
    (\ref{qSch}) with $\Delta = \gamma$. Consequently, the asymptotic can be
    static only if $\gamma\leq 1$, whereas for $\gamma >1$ the large $r$
    asymptotic can be only cosmological (KS type), and there is a horizon
    separating such an outer region from the static monopole core.

    On the other hand, if a configuration with $\gamma <1$ possesses a
    horizon, there is again a KS cosmology outside it, but there
    cannot be a large $r$ asymptotic and, according to \sect 3, the
    solutions belong to classes (b1) or (c1).

    Now, leaving aside the rather well studied case of solutions with a
    static asymptotic
\cite{Vilenkin and Shellard, Barriola and Vilenkin, Harari and Lousto},
    belonging to class (a0) according to \sect 3, let us suppose that there
    is a horizon and return to \eqs (\ref{Eq-f})--(\ref{Eq-Fom}). The
    horizon corresponds to $u\to +\infty$. For such cases, in addition to
    (\ref{bound_c}), we impose the boundary condition
\beq
    f(u) \to f_h, \cm |f_h| < \infty \cm {\rm as}
            \quad   u \to +\infty.             \label{bound_h}
\eeq
    This condition is necessary for regularity of a solution on the
    horizon and is applicable to classes (a1), (b1), (c1).

    For class (a0) solutions, having a spatial asymptotic and no horizon,
    the condition (\ref{bound_h}) is meaningless. Moreover,
    the coordinate $u$ then ranges from $-\infty$ to some $u_0 < \infty$
    such that $r (u_0) = \infty$.

    For configurations of classes (a0) and (a1), the commonly used boundary
    condition at large $r$ is
\beq
      f\to 1 \cm {\rm as} \quad r\to\infty.               \label{bound_r}
\eeq
    It is of interest that in the case (a1), to which both condition
    are applicable, the condition (\ref{bound_h}), being less restrictive,
    still leads to solutions satisfying (\ref{bound_r}) due to the
    properties of the physical system itself.

    The set of equations (\ref{Eq-f})--(\ref{Eq-Fom}) with the boundary
    conditions (\ref{bound_c}) and (\ref{bound_h}) comprise a well-posed
    nonlinear eigenvalue problem. Its trivial solution, with $f=0$ and the
    de Sitter metric (\ref{deS}), describes the symmetric state (with
    unbroken symmetry). Nontrivial solutions, describing hedgehog
    configurations with spontaneously broken symmetry, can be found
    numerically and yield a sequence of eigenvalues $\gamma_n$, $n=0,\ 1,\
    \ldots$ and the corresponding values of the horizon radius $r_{h,n}$
    for each given value of $f_h$. Conversely, for given (admissible) value
    of $\gamma$ one obtains a sequence of values of $f_h$ and $r_h$.

\subsection{Linear eigenvalue problem}

    Liebling has found empirically the upper critical value
    $\gamma_{0}\approx 3 $ for the existence of static
    solutions \cite{Liebling}%
\footnote{In the notations of \Ref{Liebling}\
      $\eta ^{\ast }\approx \sqrt{3/(8\pi)}.$}.
    In this section we find a theoretical ground for this limit. Actually,
    we find analytically a sequence of critical values $\gamma _{n},$ $n=0,
    1,...$, such that for $\gamma < \gamma_n$ there exist static
    configurations with the field magnitude $f(u)$ changing its sign $n$
    times.

\begin{figure}\centering
    \includegraphics{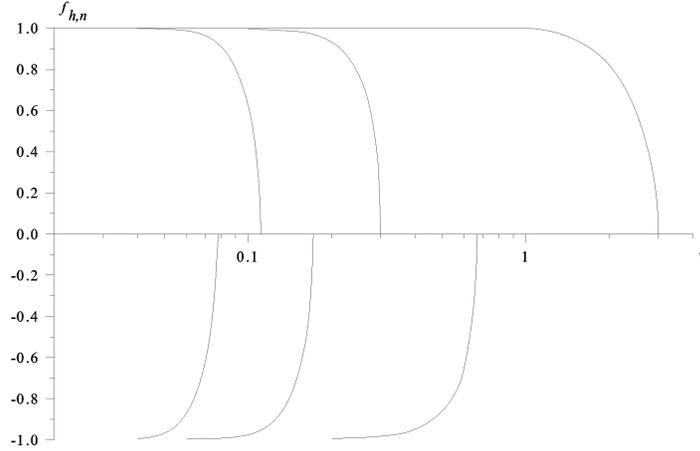}
    \caption{\label{fig:Figure3} $\gamma$ dependence of the values of $f(u)$ on the horizon,
      $f_{h,n}= f_n(u)\biggl|_{u \to\infty }$.}
\end{figure}

    In the literature one can find only an analysis for $f(u)>0$. Our
    numerical integration of \eqs (\ref{Eq-f})--(\ref{Eq-Fom}) shows that,
    in addition to solutions with monotonically growing
    $f(u)$ (which exist for $\gamma < \gamma _{0}=3$, see \fig 2a),
    for $\gamma <\gamma _{1}=2/3$ \ there are also regular solutions with
    $f(u)$ changing sign once, see \fig 2b. For $\gamma <\gamma_2=0.3$
    there are solutions with two zeros of $f(u)$ (\fig 2c), etc. All these
    solutions have a horizon, and the absolute value of $f$
    on the horizon $ |f_{h,n}| = |f_{n} (\infty )|$ is a decreasing function
    of $\gamma$, vanishing as $\gamma  \to \gamma_n-0$, see \fig 3.

    As $\gamma  \to \gamma _{n}$, the function $f (u)$ vanishes in the
    whole range of $u$, and it is this circumstance that allows us to find
    the critical values $\gamma_{n}$ analytically. In a close neighborhood
    of $\gamma _{n}$ the field $f(u)$ within the horizon is small,
    $f^{2}\ll 1$, so that \eq ({\ref{Eq-f}}) reduces to a linear
    equation with given background functions $F_{0}$ and $F_{\Omega }$,
    corresponding to the de Sitter metric (\ref{deS}). In terms of
    the dimensionless spherical radius $r$, \eq (\ref{Eq-f}) takes the form
\beq
    \frac{d}{dr} \biggl[ r^{2} \biggl( 1-\frac{r^{2}}{r_{h}^{2}}
        \biggr)\frac{df}{dr} \biggr] -  (2-r^{2}) f=0,        \label{Eq-fr}
\eeq
    where $r_h = \sqrt{12/\gamma }$ is the value of $r$
    on the horizon. The boundary conditions are
\beq
    f\Big|_{r=0} =0, \cm  |f(r_h)| < \infty.          \label{bound_lin}
\eeq

    Nontrivial solutions of (\ref{Eq-fr}) with these boundary conditions
    exist for a sequence of eigenvalues $\gamma =\gamma _{n}$,
    $n=0,1,2,...,$ and the corresponding eigenfunctions $f_{n} ( r ) $,
    regular in the interval $0\leq r\leq r_{h}$, are simple polynomials:
\beq
    f_{n} ( r ) =\sum_{k=0}^{n}a_{k} \biggl(
               \frac{r}{r_{h}} \biggr) ^{2k+1}.             \label{f_n}
\eeq
    Substituting (\ref{f_n}) into ({\ref{Eq-fr}), we find the
    eigenvalues
\beq
    r_{h,n}^{2}=2 ( 2n+1 )  (n+2 ) ,\qquad
    \gamma _{n}=\frac{3}{ ( n+1/2 )  (n+2 ) },          \label{gam_n}
\eeq
    and the recurrent relation
\beq
    a_{k}=a_{k-1}\frac{ ( 2k-1 )  ( 2k+2 )
    -r_{h,n}^{2}}{ ( 2k+1 )  ( 2k+2 )
                      -2},\qquad k=1,2,..., \label{recur}
\eeq
    allowing one to express all $a_{k},$ $k=1,2,...,n,$\ \ in terms of
    $a_{0}$. \eq ({\ref{Eq-fr}) is linear and homogeneous, so $a_{0} $ is an
    arbitrary constant%
\footnote{At $\gamma \to \gamma_n-0$, the general
       equation (\ref {Eq-f}) has the same solution as (\ref{Eq-fr}),
       with $a_0 \ll 1$. To find the dependence $a_0 (\gamma)$ one has to
       take into account the next terms nonlinear in $f$.}.
    For fixed $n$ the coefficients $a_{k}$ in  (\ref{f_n})  are
\beq
    a_{k}=a_{0}\prod_{i=1}^{k}\frac{ ( 2i-1 )  (2i+2 )
    -r_{h,n}^{2}}{ ( 2i+1 )  (2i+2 ) -2},\qquad n>0,
            \qquad 1\leq k\leq n.           \label{a_k}
\eeq

    The case
\[
    n=0,\qquad r_{h,0}=2,\qquad f_{0} (r) =a_{0}r/r_{h,0}
\]
    gives a monotonically growing function $f (u) $ in a
    close vicinity of $\gamma  = \gamma _{0}=3$, see \fig 2a.
    Thus the upper limit $\gamma _{0}=3$ for the existence of static
    monopole solutions, previously found numerically by
    Liebling \cite{Liebling}, is now obtained analytically.

    The case
\[
    n=1,        \cm       r_{h,1}=3\sqrt{2},      \cm
    f_{1} (r )
        =a_{0}\frac{r}{r_{h,{ 1}}} \biggl[ 1-\frac{7}{5}
    \biggr (\frac{r}{r_{h, 1}} \biggr) ^{2} \biggr]
\]
    describes the function $f(u)$, changing its sign once,
    at $\gamma$ close to $\gamma _{1}=2/3$, see \fig 2b. The case
    $n=2,$ $\gamma_2=3/10$, $r_{h,2}=2\sqrt{10}$,
\[
    f_{2} ( r ) =a_{0}\frac{r}{r_{h,{ 2}}} \biggl[
        1-\frac{18}{5} \biggl( \frac{r}{r_{h,{ 2}}} \biggr)
            ^{2}+\frac{99}{35}
            \biggl( \frac{r}{r_{h,{ 2}}} \biggr)^{4} \biggr]
\]
    gives the field function $f ( u ) $ changing its sign twice   (\fig 2c).

    For $n\gg 1$ the function $f_{n} ( r ) $ rapidly oscillates:
\[
    f_{n} ( r ) =a_{0}
\frac{\cos
    \biggl[ r_{h,n}\arcsin \sqrt{\fracd{r^{2}-2}{r_{h,n}^2-2}}
    -\sqrt{2}\arcsin  \biggl(\fracd{\sqrt{2}}{r}
            \sqrt{\fracd{1- ( r/r_{h,n} )^2}{1-2/r_{h,n}^{2}}}\biggl)\biggl]
  }
    {\sqrt[4]{r^2 ( r^2-2 )  \bigl[ 1- ( r/r_{h,n}) ^2 \bigr] }
    }.
\]
    However, this semiclassical formula is not valid near the left turning
    point%
\footnote{Recall that in view of the substitution (\ref{dim-0}) the
        distances are measured in the units $(\sqrt{\lambda}\eta)^{-1}$.}
    $r=\sqrt{2}$, see dashed curve on \fig 4.
    Its applicability range is $1\ll r<$ $r_{h,n}\approx 2n$, $n\gg 1$.

    We have not met so far regular monopole configurations with the field
    function $f(u) $ changing its sign. It seems that this is their first
    presentation.

\begin{figure}\centering
    \includegraphics{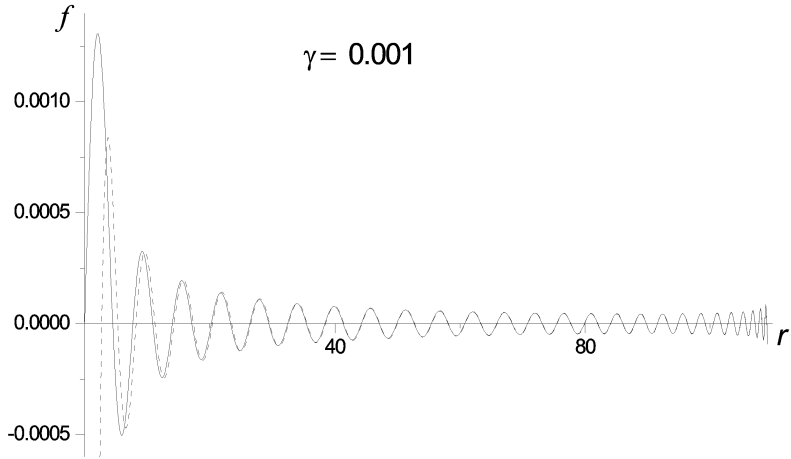}
    \caption{\label{fig:Figure4}  The field magnitude $f$ as a function of the spherical radius
      $r$ for $\gamma =0.001.$}
\end{figure}

\subsection{Solutions with monotonically growing $f(u)$}

    As is clear from the aforesaid, the interval $0<\gamma <3$ of
    existence of nontrivial solutions with monotonically growing
    $f(u) $ splits into two qualitatively different regions, separated by
    $\gamma =1$.

    In the interval $0<\gamma <1$ the solutions have the spatial
    asymptotic (\ref{qSch}) and, according to our general
    classification, belong to class (a0). The spherical radius $r(u)
    =\e^{F_{\Omega} ( u ) }$ varies from zero to infinity,
    $f ( u ) $ grows from zero to unity, $A ( u ) $
    decreases from unity to its limiting positive value (compare with
    (\ref{qSch}))
\beq                                        \label{A_infty}
    A\biggl|_{r \to \infty} =
            \frac{1-\gamma }{\alpha ^{2}},
\cm
    \alpha  =\frac{dr}{d\rho } \biggl|_{\rho  \to \infty}
    =1-\frac{\gamma}{2}\int_0^{\infty }f'{}^2(\rho) r (\rho) d\rho ,
\eeq
    and the energy integral (\ref{E}) diverges.

    In the interval $1<\gamma <3$, the solutions with monotonically
    growing $f ( u ) $ belong to class (a1).
    Instead of a spatial asymptotic, there is a horizon
    and a KS cosmology outside it. The functions $A$ and $r$ inside and
    outside the horizon are presented in \fig 5 for $\gamma=2$.

\begin{figure}\centering
    \includegraphics{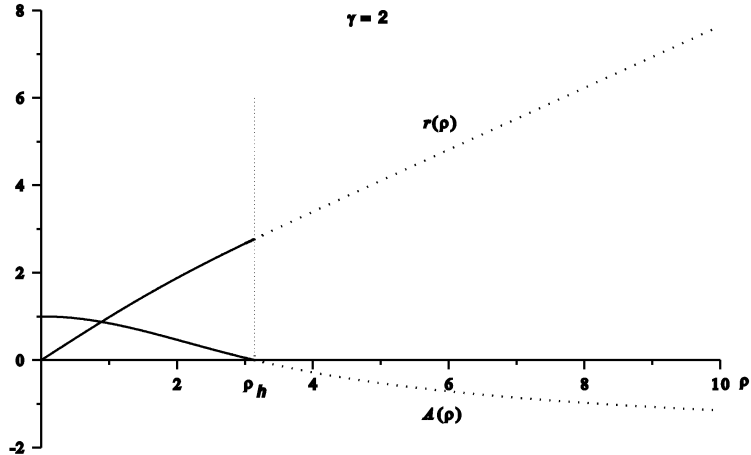}
    \caption{\label{fig:Figure5}  The functions $A( \rho ) $ and $r( \rho) $ form unified smooth
      curves in the regions inside (solid curves) and outside (dotted) the
      horizon; $\gamma =2.$}
\end{figure}

\begin{figure}\centering
    \includegraphics{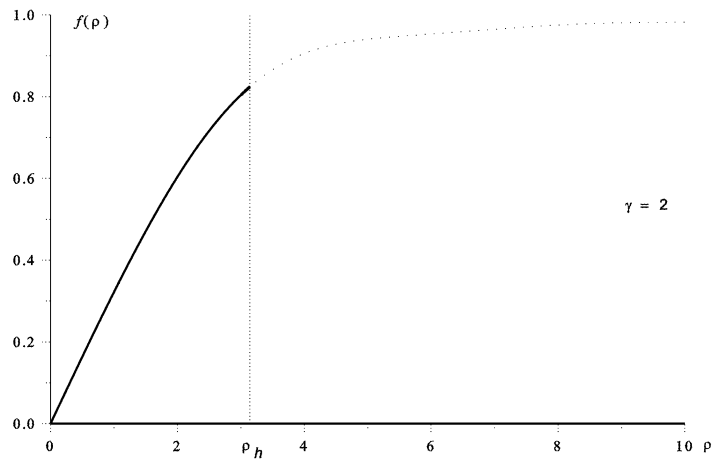}
    \caption{\label{fig:Figure6}  $f( \rho ) $ inside (solid) and outside (dotted) the horizon;
      $\gamma =2$.}
\end{figure}

    In the presence of a global monopole the cosmological expansion is
    slower than the de Sitter one (\ref{deS-tau}). As $\tau\to \infty $, the
    radius $r (\tau) $ grows linearly, while $A$ tends to the negative
    constant value $-(\gamma -1)/\alpha ^2$.

    Within the horizon $f(u)$ monotonically grows from zero at $u=-\infty $
    to a value $f_h =f_{h,0} ( \gamma ) $ on the horizon, $u \to \infty$,
    see \fig 2a. The $f$ value on the horizon $f_{h,0}$ as a function of
    $\gamma$ decreases from unity at $\gamma =1$  to  zero at $\gamma
    =\gamma _0 =3$, see \fig 3. The integral (\ref{E}) taken over
    the static region converges, and we can conclude that at $1<\gamma <3$
    the gravitational field is strong enough to suppress the Goldstone
    divergence and to localize the monopole. At $\gamma >3$ gravity
    is probably so strong that it restores the high symmetry of the system.

    Outside the horizon the field $f$\ as a function of the
    proper time $\tau $ grows from $f_{h,0}$ on the horizon
    to unity at $\tau  \to\infty$. Introducing the proper radial length $l$
    inside the horizon by the relation $dl = d\rho/\sqrt{A}$, one can
    ascertain that the functions $f(l (\rho)) $ at $\rho <h$ and
    $f (\tau( \rho  )) $ at $\rho >h$ are two parts of a single smooth
    curve, see \fig 6.

    When the parameter $\gamma$ is close to its critical value ($\gamma=1$),
    separating the (a0) and (a1) branches of the solution, i.e., for
\beq
    0 < \gamma -1 \ll 1.                            \label{gam_crit}
\eeq
    one can find analytically the horizon radius $r_{h,0}$ and
    the scalar field value on the horizon $f_{h,0}$ under certain additional
    assumptions on the system behavior which follow from the results of
    numerical analysis. In particular, there is an ``intermediate'' region
    of the $u$ range, $1 \ll u \ll u_0 = \const$, where the first term $f''$
    in the scalar field equation (\ref{Eq-f}) is very small whereas the
    function $\e^{2(F_0+F_\Omega)}$ is quite large (despite the fact that
    this function eventually vanishes as $u\to\infty$). Therefore in this
    region the expression in square brackets should be small, i.e.,
\[
    \e^{2F_\Omega } (1-f^2) \approx 2,
\]
    and this relation can be used for further estimates.

    The results are
\bearr
     \ln r_{h,0} \approx \ln [1/(\gamma-1)] \gg 1,\cm
        f_{h,0} \approx 1 - C(\gamma-1)^2,                \label{f,r_crit}
\ear
    where the constant $C$ can be found by comparison with the numerical
    results; our estimate is $C\approx 0.3$

    The solution behavior in the critical regime, $\gamma=1$, can be
    characterized as a globally static model with a ``horizon at infinity''
    \cite{Liebling} since $A\to 0$ as $r\to \infty$.

    The fact that monotonic solutions with horizons are absent for
    $\gamma <1$ becomes clear from an analysis of the inflection
    point $u=u_{\rm inf}$ of the function $F_{\Omega} (u)$.
    A horizon, if any, corresponds to $u\to \infty$ where
    $F_{\Omega }$ remains finite. Since it behaves logarithmically as
    $u\to -\infty$, there is (at least one) inflection point,
    where the second-order derivative is zero, and from
    \eq (\ref{Eq-Fom}) we have
\[
    1-\gamma f^2-\frac{\gamma }{4}\e^{2F_\Omega} (f^2-1)^2=0,
             \cm    u=u_{\rm inf}.
\]
    This is a quadratic equation with respect to $1-f^{2}$, whence
\beq                                                         \label{inflec}
    1-f^{2} = 2 \e^{-2F_{\Omega }} \biggl(1 \pm
        \sqrt{1-\frac{\gamma -1}{\gamma}\e^{2F_\Omega}} \biggr).
\eeq
    A monotonically growing function $f(u)$ corresponds to greater
    values of $f$, i.e., to the ``minus'' branch of (\ref{inflec}) (as is
    confirmed by numerical results). But then the r.h.s. of (\ref{inflec})
    is negative for $\gamma < 1$, leading to $f^2 >1$, which cannot happen
    since $f^2=1$ is the maximum attainable value for the solutions under
    study.  Thus, for $\gamma<1$, all solutions with monotonically growing
    $f(u)$ belong to class (a0), possess a spatial asymptotic with a solid
    angle deficit and a divergent field energy.

    Numerical integration confirms these conclusions.
    The different behavior of $F_{\Omega} (u)$ for $\gamma <1 $
    and $\gamma >1$ is shown in  \fig 7.

\begin{figure}\centering
    \includegraphics{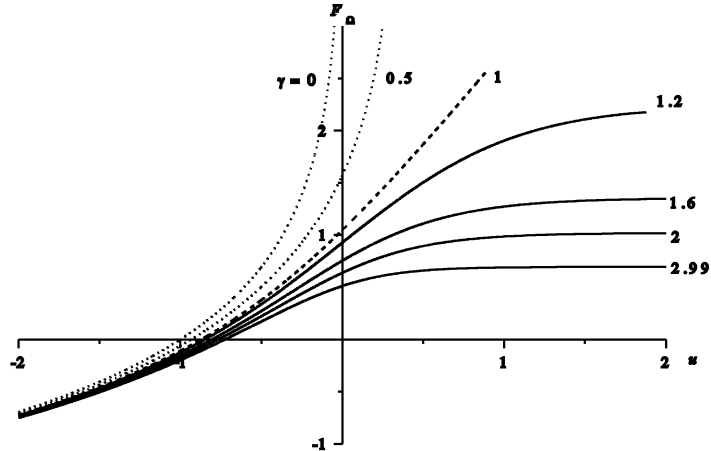}
    \caption{\label{fig:Figure7}  The function $F_{\Omega}( u) $ for different
     values of $\gamma$. At $\gamma <1$ there is a limiting value
     $u_{\max}$ of $u$, such that $F_{\Omega }\to \infty$ as
     $u \to u_{\max}(\gamma )$ (dotted curves). As $\gamma \to 1$, the value $u_{\max}(\gamma)
     \to \infty$ (dashed curve), and, for $\gamma >1$, $F_{\Omega }( u) $ tends to a
     finite constant value as $u\to \infty$ (solid curves).}
\end{figure}

\subsection{Solutions with $f(u)$ changing its sign}

    For $\gamma <\gamma_1=2/3$ there are solutions with the function
    $f(\phi)$ changing its sign once, see \fig 2b. For $\gamma <\gamma
    _{2}=0.3$ there are solutions where $f(\phi)$ changes its sign
    twice, see \fig 2c, etc. Unlike the monotonic solutions discussed in
    \sect 4.3, all of them possess a horizon, and, in agreement with
    the general inferences of \sect 4.1, they belong to class (c1).
    This means that, beginning with a regular center, the spherical radius
    $r(\rho)$ first grows, then passes its maximum $r_{\max}$ at some
    $\rho _1$ and then decreases to zero at finite $\rho =\rho_2$ which is a
    singularity. The horizon occurs at some $\rho =h < \rho_2$, which can be
    greater or smaller than $\rho_1$, but in any case the singularity
    takes place in a T-region and is of cosmological nature.
    The dependence $r (\rho) $ before and after the horizon is a single
    smooth curve, see \fig 8a.

\begin{figure}\centering
    \includegraphics{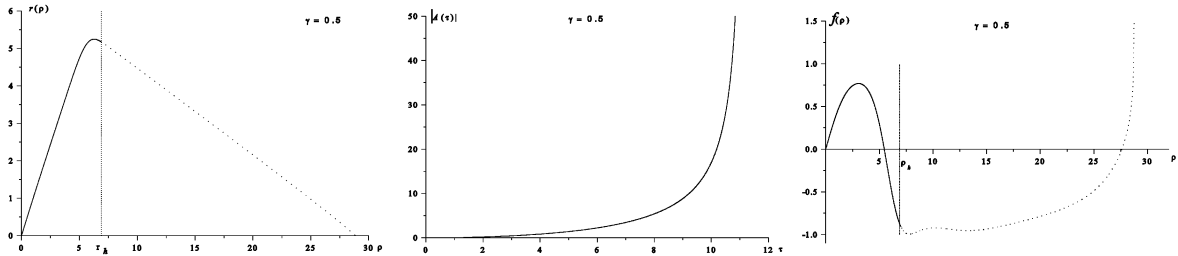}
    \caption{\label{fig:Figure8}  Solutions with $f $ of alternating sign:
      the functions (a) $r( \rho ), $ (b) $|A( \rho )|, $ and (c) $f( \rho ) $ for $\gamma =0.5$
      before (solid) and after (dotted) the horizon.}
\end{figure}

    Beyond the horizon, $ |A (\tau)| $ as a function of the proper time
    $\tau$ of a comoving observer grows from zero at $\tau =0$ (the horizon)
    to infinity at $\tau  \to \tau _{s}=\tau  (\rho_2)$ (the singularity),
    see \fig 8b. The scalar field magnitude $| f(\tau) |$ beyond the horizon
    first grows, then slightly varies around unity. Approaching the
    singularity, $f (\tau) $ changes its sign and finally $| f (\tau)| \to
    \infty $ as $\tau  \to \tau_{s}$, see \fig 8c.

\section{Conclusion and discussion}

    We have performed a general study of the properties
    of static global monopoles in general relativity. We have shown that,
    independently of the shape of the symmetry breaking potential, the
    metric can contain either no horizon, or
    one simple horizon, and in the latter case the space-time global
    structure is the same as that of de Sitter space-time. Outside
    the horizon the geometry corresponds to homogeneous
    anisotropic cosmological models of KS type, where
    spatial sections have the topology $\R\times \S^2$. In
    general, all possible solutions can be divided into six classes with
    different qualitative behavior. This classification is obtained
    without any assumptions about $V (\phi) $. Solutions
    with given $V (\phi) $ contain some of these
    classes, not necessarily all of them. This qualitative analysis
    gives a complete picture of what can be expected for global
    monopole systems with particular symmetry breaking potentials.

    Our analytical and numerical analysis for the particular case of
    ``Mexican hat'' potential confirmed the previous results of other
    authors concerning the configurations with monotonically growing
    order parameter. Among other things, we have obtained analytically
    the upper limit $\gamma _0=3$ for the existence of static
    monopole solutions, previously found numerically by
    Liebling \cite{Liebling}. We have also found and analyzed a new
    family of solutions with the field function $f$ changing its sign, which
    we have not met in the existing literature.

    Of particular interest can be the class (a1) solutions with a static
    nonsingular monopole core and a KS cosmological model outside the
    horizon. Its anisotropic evolution is determined by the functions
    of the proper time $|A (\tau)|$ (the squared scale factor in the $\R$
    direction, $r (\tau)$ (the scale factor in the two $\S^2$ directions)
    and the field magnitude $f(\tau)$. For a comoving observer in the
    T-region, the expansion starts with a rapid growth of $|A(\tau)|$
    from zero to finite values, resembling inflation, and ending with
    $A\to \const$ as $\tau\to\infty$. The expansion in the $\S^2$
    directions, described by $r(\tau)$, is comparatively uniform and linear
    at late times, i.e., much slower than the de Sitter's
    $\cosh^2(\tau /r_h)$, see (\ref{deS}). It should be stressed that all
    such models with de Sitter-like causal structure, i.e., a static core
    and expansion beyond a horizon, drastically differ from standard
    Big Bang models in that the expansion starts from a nonsingular
    surface, and cosmological comoving observers can receive information in
    the form of particles and light quanta from the static region, situated
    in the absolute past with respect to them. Moreover, in our case the
    static core is nonsingular, and it is thus an example of an entirely
    nonsingular cosmology in the spirit of papers by Gliner and Dymnikova
    \cite{Gliner,Dymnikova}.

    The nonzero symmetry-breaking potential plays the role of a
    time-dependent cosmological constant, a kind of hidden vacuum matter.
    Since the field function $f$ tends to unity as $\tau  \to \infty$,
    the potential vanishes, and the ``hidden vacuum matter'' disappears.

    The lack of isotropization at late times does not seem to be a fatal
    shortcoming of the model for two reasons. First, if the model is
    applied for describing the near-Planck epoch of the Universe evolution,
    then, on the next stage, the anisotropy can probably be damped by
    diverse particle creation, and the further stages with lower energy
    densities may conform to the standard picture (with possible further
    phase transitions). Second, if we add a comparatively small positive
    quantity $\Lambda$ to the potential (\ref{hat}) (``slightly raise the
    Mexican hat''), this must change nothing but the late-time asymptotic
    which will become de Sitter, corresponding to the cosmological constant
    $\Lambda$. In our view, these ideas deserve a further study.

    Evidently, the present simple model cannot be directly applied to our
    Universe. It would be too naive to expect that a macroscopic description
    based on a simple toy model of a global monopole with only one
    dimensionless parameter $\gamma $ can explain the whole variety of
    early-Universe phenomena. Nevertheless, it may be considered as an
    argument in favor of the idea that the standard Big Bang might be
    replaced with a nonsingular static core and a horizon appearing as a
    result of some
 symmetry-breaking phase transition on the Planck
    energy scale.

\section{Acknowledgement}

The authors are grateful to Acad. A.F. Andreev for a useful
discussion at the seminar at P.L. Kapitza Institute of
Physical Problems.

\end{document}